# Hydrostatic pressure induced transition from $\delta T_c$ to $\delta\ell$ pinning mechanism in MgB$_2$

Babar Shabbir[1], Xiaolin Wang[1], S. R. Ghorbani[1,2], and Shixue Dou[1].

[1]*Spintronic and Electronic Materials Group, Institute for Superconducting & Electronic Materials, Australian Institute for Innovative Materials, University of Wollongong, Wollongong, NSW, 2500, Australia*

[2]*Department of Physics, Ferdowsi University of Mashhad, Mashhad, 9177948974, Iran*

The impact of hydrostatic pressure up to 1.2 GPa on the critical current density (J$_c$) and the nature of the pinning mechanism in MgB$_2$ have been investigated within the framework of the collective theory. We found that the hydrostatic pressure can induce a transition from the regime where pinning is controlled by spatial variation in the critical transition temperature ($\delta T_c$) to the regime controlled by spatial variation in the mean free path ($\delta\ell$). Furthermore, T$_c$ and low field J$_c$ are slightly reduced, although the J$_c$ drops more quickly at high fields than at ambient pressure. We found that the pressure raises the anisotropy and reduces the coherence length, resulting in weak interaction of the vortex cores with the pinning centres. Moreover, the hydrostatic pressure can reduce the density of states [N$_s$(E)], which, in turn, leads to a reduction in the critical temperature from 39.7 K at P = 0 GPa to 37.7 K at P = 1.2 GPa.

Magnesium diboride (MgB$_2$) is a promising superconducting material which can replace conventional low critical temperature (T$_c$) superconductors in practical applications, due to its relatively high T$_c$ of 39 K, strongly linked grains, rich multiple band structure, low fabrication cost, and especially, its high critical current density (J$_c$) values of $10^5$-$10^6$ A/cm$^2$ [1-9]. Numerous studies have been carried out in order to understand the vortex-pinning mechanisms in more detail, which have led to real progress regarding the improvement of J$_c$. There are two predominant mechanisms, $\delta T_c$ pinning, which is associated with spatial fluctuations of the T$_c$, and $\delta\ell$ pinning, associated with charge carrier mean free path ($\ell$) fluctuations [10-14].

Very recently, our team have found that hydrostatic pressure is a most effective approach to enhance J$_c$ significantly in iron based superconductors, as the pressure can induce more point pinning centres and also affect the pinning mechanism [15]. Therefore, it is natural to investigate the impact of hydrostatic pressure on J$_c$ and flux pinning mechanisms in MgB$_2$. Previous studies have shown that pressure of 1 GPa can reduce T$_c$, but only by less than 2 K in MgB$_2$. This is a very insignificant reduction as compared to the other approaches (i.e. chemical doping and irradiation) which are mainly used for J$_c$ enhancement [16]. For instance, chemical doping can significantly enhance J$_c$ in MgB$_2$, but with a considerable degradation of T$_c$; carbon doping can reduce T$_c$ from 39 K to nearly as low as 10 K, for carbon content up to 20% [17-22]. Similarly, the irradiation method can improve J$_c$ in MgB$_2$, but it reduces T$_c$ values significantly (by more than 20 K in some cases) [23-27]. Correspondingly, the chemical doping and irradiation methods can also change the nature of the pinning mechanism in MgB$_2$ [24, 28-30]. The determination of J$_c$ and the flux pinning mechanism under hydrostatic pressure is also an important step to probe the mechanism of superconductivity in more detail in MgB$_2$. It is very interesting to know whether hydrostatic pressure can increase the pinning and J$_c$ at both low and high fields.

In this work, we report our study on pressure effects on T$_c$, J$_c$, and the flux pinning mechanism in MgB$_2$. Hydrostatic pressure can induce a transition from the regime where pinning is controlled by spatial variation in the critical transition temperature ($\delta T_c$) to the regime controlled by spatial variation in the mean free path ($\delta\ell$). In addition, T$_c$ and low field J$_c$ are slightly reduced, although the J$_c$ drops more quickly at high fields than at ambient pressure. We found that the pressure increases the anisotropy and reduces the coherence length, resulting in weak interaction of the vortex cores with the pinning centres.

The MgB$_2$ bulk sample used in the present work was prepared by the diffusion method. Firstly, crystalline boron powders (99.999%) with particle size of 0.2-2.4 μm were pressed into pellets. They were then put into iron tubes filled with Mg powder (325 mesh, 99%), and the iron tubes were sealed at both ends. Allowing for the loss of Mg during sintering, the atomic ratio between Mg and B was 1.2:2. The sample was sintered at 800°C for 10 h in a quartz tube under flowing high purity argon gas. Then, the sample was furnace cooled to room temperature. The temperature dependence of the magnetization and the M-H loops at different temperatures and pressures were performed on QD PPMS (14T) by using vibration sample magnetometer (VSM). We used an HMD High Pressure cell with Daphne 7373 oil as a pressure transmission medium to apply hydrostatic pressure on a sample. The critical current density was calculated by using the Bean approximation.

The zero-field-cooling (ZFC) and field-cooling (FC) curves at different applied pressures are plotted in Fig. 1. The T$_c$ drops from 39.7 K at $P = 0$ GPa to 37.7 K at $P = 1.2$ GPa, with a pressure coefficient of -1.37 K/GPa, as can be seen in the inset of Fig. 1. It is well known that T$_c$, the unit cell volume (V), and the anisotropy (γ) under pressure can be interrelated through a mathematical relation as in [33]

$$\Delta T_c^/(P) + \Delta V^/ + \Delta\gamma^/ = 0 \qquad (1)$$

where

$$\Delta T_c^/(P) = \left[\frac{T_c(P)-T_c(0)}{T_c(0)}\right], \quad \Delta V^/ = \left[\frac{V(P)-V(0)}{V(0)}\right] \quad \text{and}$$
$$\Delta \gamma^/ = \left[\frac{\gamma(P)-\gamma(0)}{\gamma(0)}\right].$$

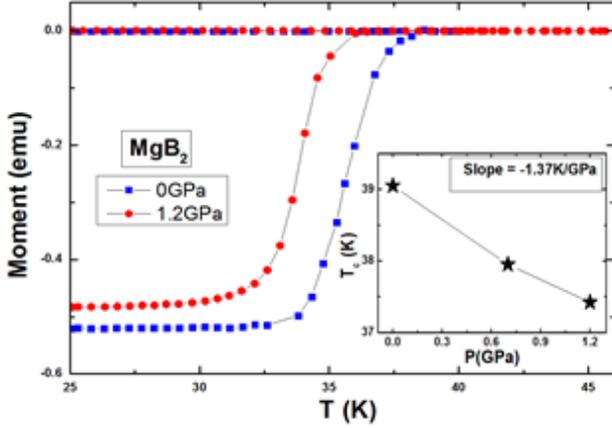

**Figure 1:** Temperature dependence of magnetic moment under different applied pressures in both ZFC and FC runs. The inset shows the pressure dependence of the critical temperature for $MgB_2$. $T_c$ is found to decrease with a slope of $dT_c/dP$ = -1.37 K/GPa.

The $\Delta V^/$ found for $MgB_2$ is 0.0065, as the pressure can reduce the unit cell volume of $MgB_2$ from 29.0391 Å$^3$ at $P$ = 0 GPa to 28.8494 Å$^3$ at $P \approx 1.2$ GPa [34]. A similar value for $\Delta V^/$ can also be obtained from $\Delta V^/ = -\Delta P/B$, where $B$ is the bulk modulus of the material [33]. We found $\Delta T_c^/(P)$ = 0.042 from Figure 1. By using $\Delta V^/$ and $\Delta T_c^/(P)$, we can obtain from Equation (1):

$$\Delta \gamma^/ = \left[\frac{\gamma(P)-\gamma(0)}{\gamma(0)}\right] \approx 0.036 \quad (2)$$

This indicates that the anisotropy of $MgB_2$ is increased by applying pressure, *i.e.*, $\gamma(P) > \gamma(0)$. Therefore, the coherence length ($\xi$) at $P$ = 1.2 GPa is reduced as compared to its value at $P$ = 0 GPa [*i.e.*, $(\xi)_P < (\xi)_0$]. The density of states in Bardeen-Cooper-Schrieffer (BCS)-like superconductors such as $MgB_2$ is expressed as

$$N_s(E) = N_n(E_F)\left[\frac{E}{\sqrt{E^2-\Delta^2}}\right] \quad (3)$$

Where $N_n(E_F)$ is the density of states at the Fermi level in the normal state and $\Delta$ is the superconductivity gap. Therefore, $N_s(E) \propto N_n(E_F)$ and

$$N_n(E_F) \propto VE_F^{1/2} \propto Vk_F^2 \quad (4)$$

where $V$ is the total volume and $k_F$ is the Fermi wave vector [35, 36],

$$k_F = \frac{2m\Delta\xi}{\hbar}. \quad (5)$$

Combining Equations (3), (4), and (5), we obtain
$$N_s(E) \propto V\xi. \quad (6)$$

It is important to mention that pressure has no significant impact on the unit cell volume of $MgB_2$ up to $P$ = 1.2 GPa. Therefore, the density of states is mainly dependent on $\xi$. $(\xi)_P < (\xi)_0$ leads to a comparison regarding the density of states at $P$ = 1.2 GPa and $P$ = 0 GPa

$$i.e. \; [N_s(E)]_P < [N_s(E)]_0, \quad (7)$$

given that hydrostatic pressure can decrease the density of states in $MgB_2$ and therefore contributes to a reduction in $T_c$.

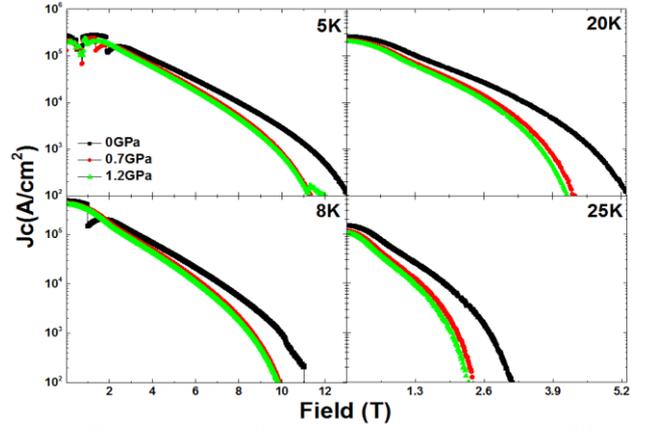

**Figure 2:** Field dependence of critical current density ($J_c$) under different pressures measured at 5 K, 8 K, 20 K, and 25 K.

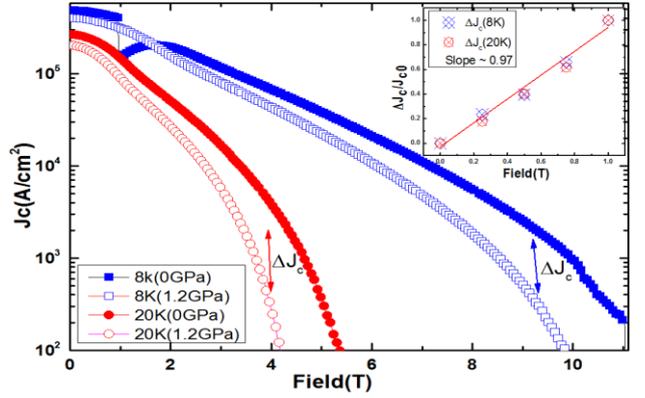

**Figure 3:** Comparison of $J_c$ at two pressures (0 GPa and 1.2 GPa) for 8 K and 20 K curves. The inset shows the plot of $\Delta J_c/J_{c0}$ versus field, representing the trend towards the suppression of $J_c$ with increasing field, nearly at a same rate of ~ -0.97 T$^{-1}$ for 8 K and 20 K.

Fig. 2 shows the field dependence of $J_c$ at different temperatures (i.e. 5, 8, 20, and 25 K) and pressures (i.e. 0, 0.7, and 1.2 GPa). We found that low field $J_c$ was reduced slightly under pressure. The $J_c$ drops more quickly at high fields, however, as compared to $P$ = 0 GPa. This is further reflected in Fig. 3, which shows $J_c$ values at 8 and 20 K under pressure. The inset shows normalized $\Delta J_c$ (*i.e.*, $\Delta J_c = J_c^P - J_c^o$) for both 8 K and 20 K, which indicates almost a similar decay trend.

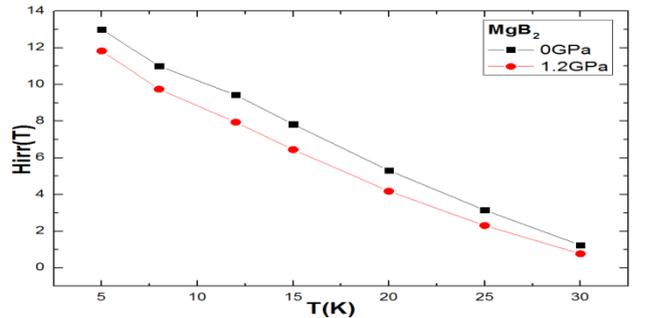

**Figure 4:** $H_{irr}$ as a function of temperature.

We also plotted $H_{irr}$ as a function of temperature in Fig. 4, which shows that $H_{irr}$ decreases gradually from nearly 13 T to 11.8 T at $T$ = 5 K for $P$ = 1.2 GPa, which is ascribed to the observed $J_c$ suppression.

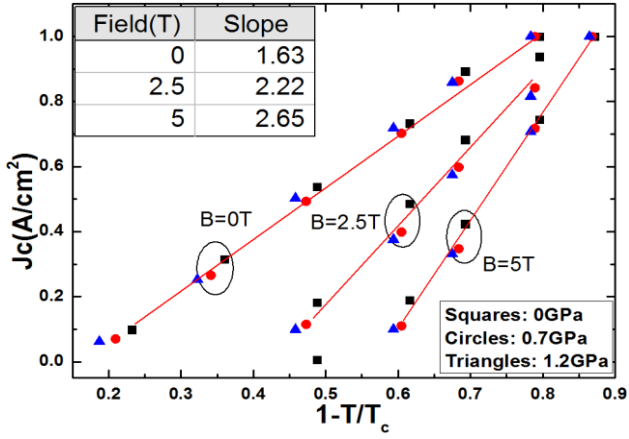

Figure 5: $J_c$ as a function of reduced temperature ($\tau = 1 - T/T_c$) at 0, 2.5, and 5 T for pressures of 0, 0.7, and 1.2 GPa. The solid lines are fitted well to the data according to the power law in the framework of Ginzburg-Landau theory.

$J_c$ as a function of reduced temperature ($\tau = 1 - T/T_c$, where $T$ is the temperature and $T_c$ is the critical temperature) is plotted in Fig. 5. The temperature dependence of $J_c$ follows a power law description in the form of $J_c \propto \tau^\mu$, where $\mu$ is the slope of the fitted line and its value depends on the magnetic field [37-39]. The exponent $\mu$ in our case is found to be nearly same at different pressures, and its values are 1.63, 2.22, and 2.65 at fields of 0, 2.5, and 5 T, respectively. Different values of exponent $\mu = 1$, 1.7, 2, and 2.5 are also reported for standard yttrium barium copper oxide (YBCO) films [40]. The larger exponent value at high field shows that pressure effects are more significant at high fields as compared to low fields.

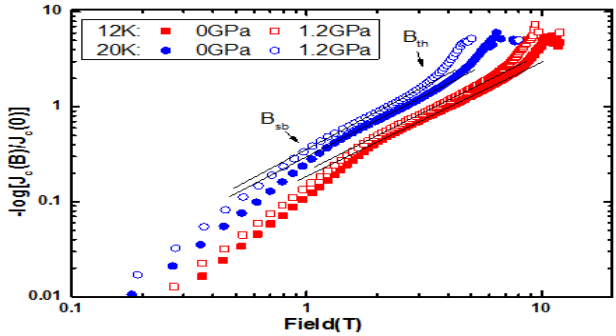

Figure 6: Double logarithmic plot of $-\log[J_c(B)/J_c(0)]$ as a function of field at 12 K and 20 K.

A double logarithmic plot of $-\log[J_c(B)/J_c(0)]$ as a function of field at 12 K and 20 K for $P = 0$ GPa and $P = 1.2$ GPa is plotted in Fig. 6. This shows deviations at certain fields, denoted as $B_{sb}$ and $B_{th}$. According to the collective theory [10], the region below $B_{sb}$ is the regime where the single-vortex-pinning mechanism governs the vortex lattice in accordance with the following expression,

$$B_{sb} \propto J_{sv} B_{c2} \quad (8)$$

Where, $J_{sv}$ is the critical current density in the single vortex pinning regime and $B_{c2}$ is the upper critical field. At high fields (above the crossover field $B_{sb}$), $J_c(B)$ follows an exponential law

$$J_c(B) \approx J_c(0) \exp\{-(B/B_0)^{3/2}\} \quad (9)$$

Where, $B_0$ represents a normalization parameter on the order of $B_{sb}$. It is well known that the deviation observed at $B_{sb}$ is linked to the crossover from the single-vortex-pinning regime to the small-bundle-pinning regime, while the deviation at the thermal crossover field ($B_{th}$) can be connected to large thermal fluctuations [8].

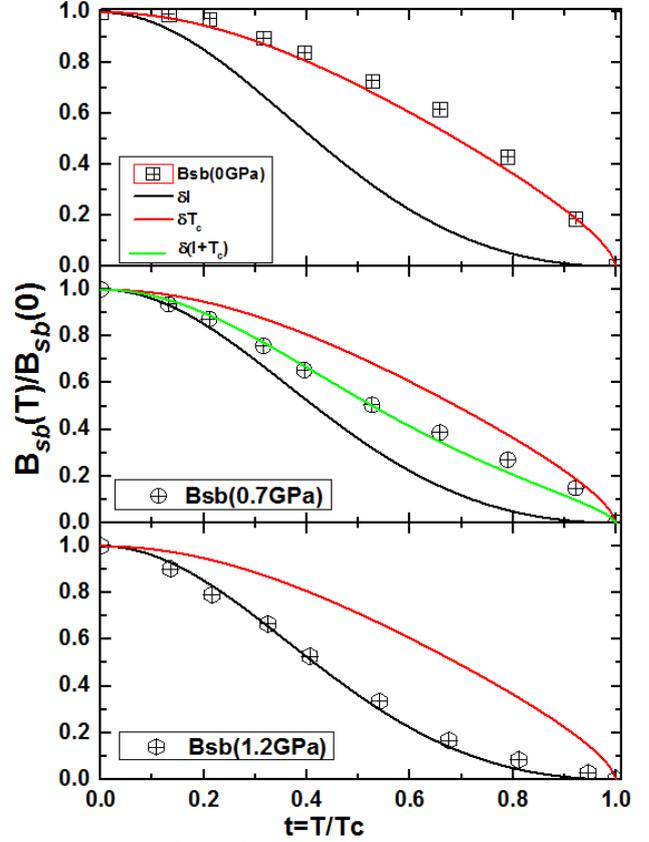

Figure 7: Plots of $B_{sb}(T)/B_{sb}(0)$ vs. $T/T_c$ at different pressures (0, 0.7, and 1.2 GPa). The red fitted line is for $\delta T_c$ pinning, the black fitted line is for $\delta \ell$ pinning, and the green fitted line is for mixed $\delta(T_c + \ell)$ pinning.

The pinning behaviour can be obtained from the temperature dependence of the crossover field from the single vortex regime [41]. The temperature dependence of the crossover field can be expressed as

$$B_{sb}(T) = B_{sb}(0)\left(\frac{1-t^2}{1+t^2}\right)^v \quad (10)$$

Where $v = 2/3$ and 2 for $\delta T_c$ and $\delta \ell$, respectively.
The above-mentioned Equation (10) can be found by inserting the following expressions with $t = T/T_c$ into Equation (8),

$$J_{sv} \approx (1-t^2)^{7/6}(1+t^2)^{5/6} \quad : \text{for } \delta T_c \quad (11)$$

and $J_{sv} \approx (1-t^2)^{5/2}(1+t^2)^{-1/2}$: for $\delta \ell \quad (12)$

The crossover fields ($B_{sb}$) for reduced temperature ($T/T_c$) at $P = 0$ GPa, 0.7 GPa, and 1.2 GPa are plotted in Fig. 7. The experimental data points for $B_{sb}$ are scaled through Eq. (10) for $\delta \ell$ and $\delta T_c$. We found that hydrostatic pressure can induce the transition from the $\delta T_c$ to the $\delta \ell$ pinning mechanism. The $\delta T_c$ pinning mechanism is dominant in pure MgB$_2$ polycrystalline bulks, thin films, and single crystals [14, 42, 43]. The coherence length is proportional to the mean free path $\ell$ of the carriers, and therefore, pressure can enhance $\delta \ell$ pinning in MgB$_2$. It is noteworthy that $J_c$ drops under pressure in MgB$_2$ due to the transition in the flux pinning mechanism.

In summary, the impact of hydrostatic pressure on the $J_c$ and the nature of the pinning mechanism in MgB$_2$, based on the collective theory, have been investigated. We found that the hydrostatic pressure can induce a transition from the $\delta T_c$ to the $\delta\ell$ pinning mechanism. Furthermore, pressure can slightly reduce low field $J_c$ and $T_c$, although pressure has a more pronounced effect on $J_c$ at high fields. Moreover, the pressure can also increase the anisotropy, along with causing reductions in the coherence length and H$_{irr}$, which, in turn, leads to a weak pinning interaction.

Acknowledgments

X.L.W. acknowledges support from the Australian Research Council (ARC) through an ARC Discovery Project (DP130102956) and an ARC Professorial Future Fellowship project (FT130100778). Dr. T. Silver's critical reading of this paper is greatly appreciated.

Author Contributions

X.L.W conceived the pressure effects and designed the experiments. B.S. performed high pressure measurements. X.L.W and B.S analysed the data and wrote the paper. All authors contributed to the discussions of the data and the paper.

Additional information

Competing financial interests: The authors declare no competing financial interests.